# Protein-Based Electrical Junctions with Robust Biocompatible Carbon Electrodes Exhibit Activation-less Charge Transport down to 10 K

Shailendra K. Saxena,[†,¥,€] Sudipta Bera,[†,¥] Tatyana Bendikov,[§] Israel Pecht,[‡] Mordechai Sheves,[†,*] and David Cahen[†,*]

[†]Department of Molecular Chemistry and Materials Science, Weizmann Institute of Science, Rehovot 7610001, Israel

[§]Department of Chemical Research Support, Weizmann Institute of Sciences, Rehovot 7610001, Israel

[‡]Department of Immunology and Regenerative Biology, Weizmann Institute of Science, Rehovot 7610001, Israel

[€]Department of Physics and Nanotechnology, College of Engineering and Technology, SRM Institute of Science and Technology, Kattankulathur, Chennai 603203, Tamil Nadu, India

[¥]The authors contributed equally to this work

[*]Corresponding Authors' email: *mudi.sheves@weizmann.ac.il*, *david.cahen@weizmann.ac.il*

**ORCiD**

Shailendra K. Saxena: 0000-0001-7156-3407
Sudipta Bera: 0000-0001-7894-9249
Tatyana Bendikov: 0000-0002-1637-6366
Israel Pecht: 0000-0002-1883-9547
Mordechai Sheves: 0000-0002-5048-8169
David Cahen: 0000-0001-8118-5446

**Abstract**

The integration of functional proteins into solid-state electronic devices remains a central challenge in molecular bioelectronics due to the fragile nature of protein structures and their complex charge-transport behaviour. Here, we present a robust crosswire evaporated top-contact device based on bacteriorhodopsin (bR) single-bilayers (SBL), configured as Au/Cys/bR(SBL)/eC/Au (simplified as Au/bR/eC). The evaporated carbon (eC) top electrode forms a conformal, non-invasive contact that suppresses filament formation and ensures electrical integrity across the cross-wire intersecting area (~200 $\mu m^2$). Structural and spectroscopic analyses confirm that the solid-state bR films maintain the native absorption spectrum and have functional photocycle activity after electrode deposition, implying that their native conformation is not significantly affected. Remarkably, electron transport (ETp) through the ~9 nm bR-SBL junctions is temperature-independent within 300 K – 10 K, excluding thermally activated hopping, while the length is incompatible with coherent tunneling. Under green illumination, the junctions exhibit a reversible, photo-induced current enhancement ($J_{green}/J_{dark} \approx 2$), ascribed to light-driven conformational changes rather than direct photoexcitation. The Au/bR/eC architecture thus establishes a thermally non-activated, conformationally mediated transport mechanism via a stable, cryo-compatible solid-state protein junction. This work provides a scalable platform for integrating light-responsive biomolecules into future bio-optoelectronic and neuromorphic devices.

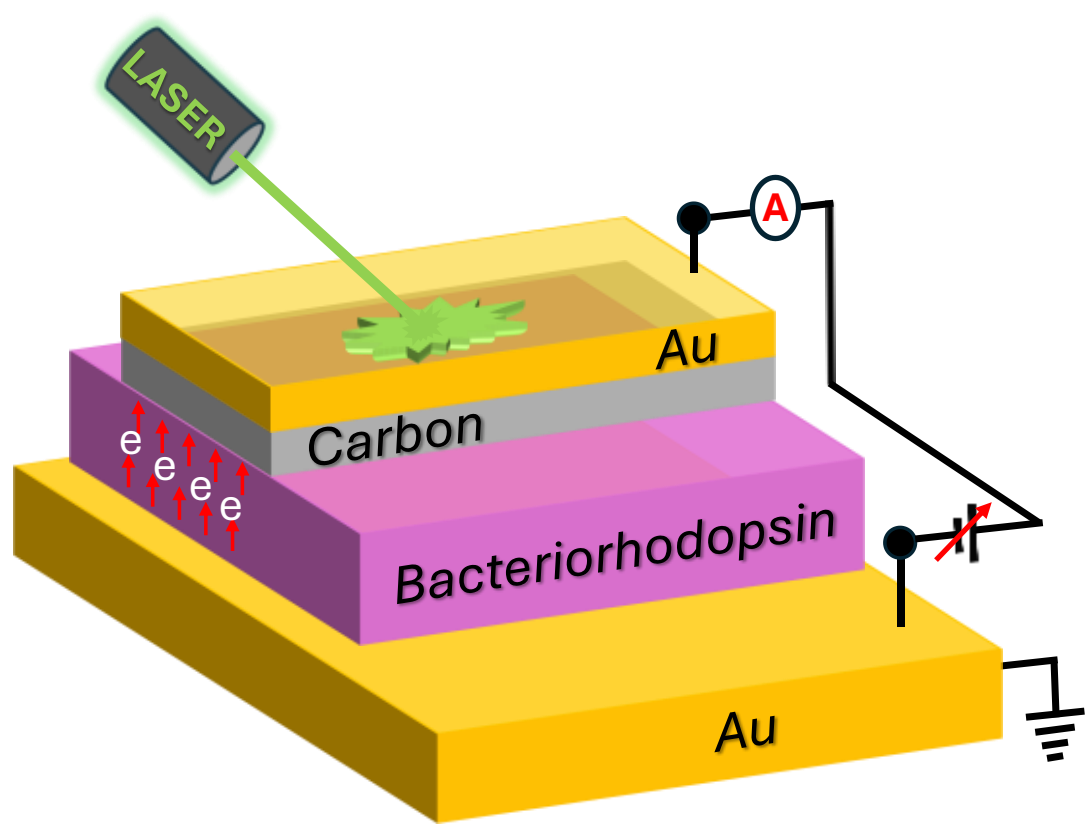

## 1. Introduction

Proteins, due to their intrinsic structural diversity and functional specificity, have emerged as exceptional building blocks for bio-compatible electronic devices. In bioelectronics,[1,2] biomolecules, particularly proteins, show promise for integration into solid-state devices as active, electronically conducting elements such as nano-scale transistors,[3] logic circuits,[4] etc. The demonstration that ultrathin protein films can sustain electron transport (ETp) highlights their potential to combine biochemical selectivity, structural self-assembly, and electronic functionality within a single material platform.[5–7] This capability paves the way for sensitive, selective, and adaptive bioelectronic interfaces, enabling direct communication with biological systems, and for innovative avenues in diagnostics, drug delivery, and personalized nanomedicine.

Charge transport through solid-state protein junctions[8] involves pure electronic conduction, without measurable ionic diffusion,[9] positioning proteins as natural conduits for nanoscale electron transport. Results of extensive studies on self-assembled protein monolayers (SAMs) have often been interpreted in terms of one of two dominant transport mechanisms: quantum tunneling[5,10] and thermally activated hopping.[11,12] However, our recent findings suggest a temperature-independent, activation-less ETp process operating over junction lengths of several tens of nanometers,[9,13] is not compatible with either tunneling or hopping. Thus, resolving the mechanism(s) controlling charge transport via protein junctions is pivotal for developing efficient bioelectronic devices.[14,15]

To address this, we recently developed a metal/protein/metal-based micropore device (MpD)[13] that provides permanent electrical contact while minimizing/ eliminating interface effects, enabling measurement of intrinsic protein electron transport properties.[16] Because fabrication of the multilayered device structure of MpD is challenging, we also sought a simpler biocompatible alternative for producing stable electrical contacts that will at least minimize contact effects.

Inspired by the work of McCreery and coworkers, who demonstrated that an evaporated carbon layer can serve as a stable, non-filamentary top electrode for molecular junctions of conjugated organic layers,[17–21] we set out to create a carbon top-electrode deposited protein junction. The reactive yet non-invasive nature of evaporated carbon permits deposition directly onto the protein film without causing an electrical shortage. Moreover, the optical transparency of ultra-thin carbon layers[22–24] enables in situ monitoring of protein photoactivity, important for verifying the functional integrity of the proteins within solid-state devices.

To be able to use photoactivity as a measure of protein functionality, we selected bR, because it is known to combine photochemical activity with electronic transport. bR functions as a well-characterized[25,26] light-driven proton pump that combines long lifetimes, robust stability, and self-assembly into ordered films,[14] making it very suitable for studying photo-responsive ETp in solid-state configurations. The organic–carbon interface reduces contact resistance [16,27] and improves charge injection, providing a transparent, biocompatible platform for reliable ETp measurements and functional bioelectronic devices.

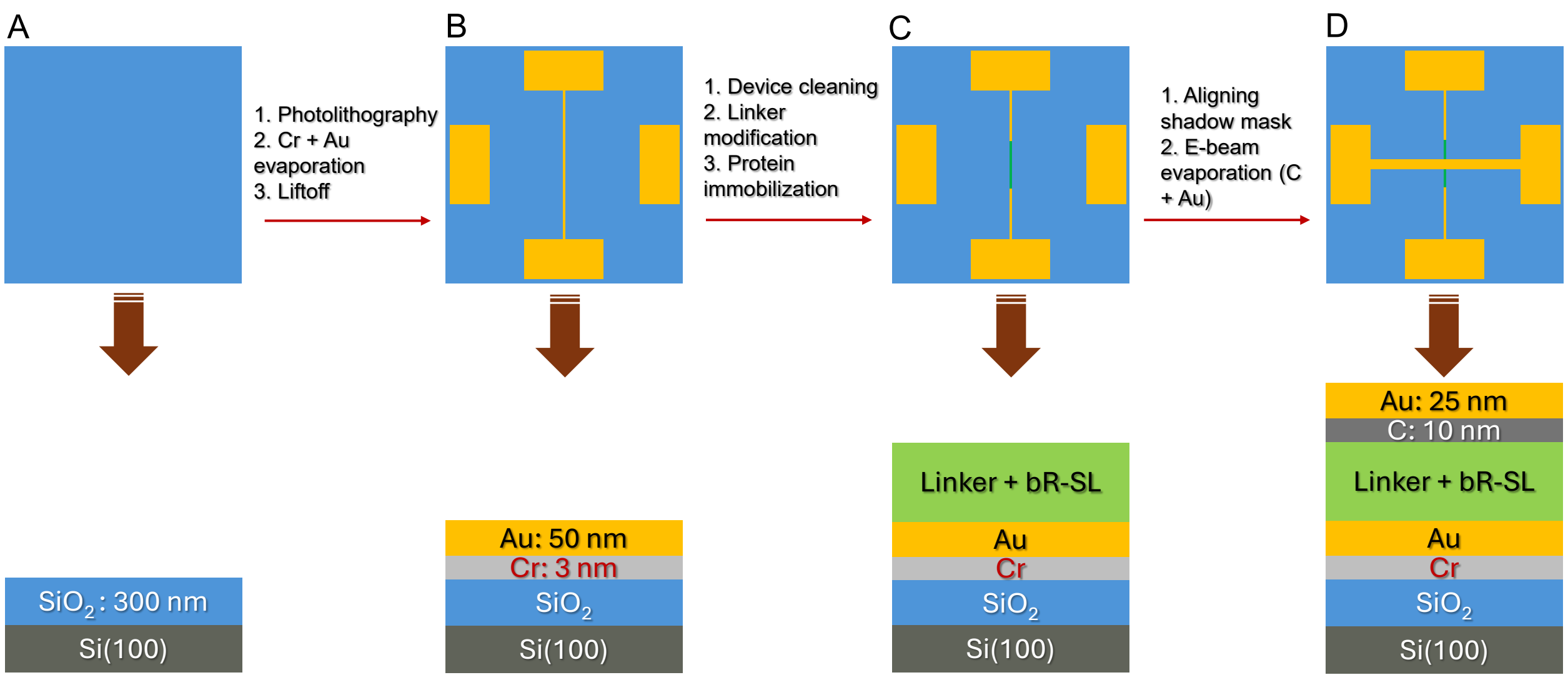

**Scheme 1.** (A–D) Top-view schemes to illustrate the step-by-step fabrication process of the Au/bR/eC crosswire device. (A) Bare $SiO_2$ – on - Si substrate. (B) Patterned bottom electrode: vertical yellow line of Au with four large Au contact pads for electrical probing. (C) Bottom electrode modified with bacteriorhodopsin (bR) protein layer by surface functionalization via linker molecules. (D) Final Au/bR/eC device structure with a cross-aligned top Au on eC electrode; the intersection of the orthogonal electrode lines defines the active protein junction region ($A_{geo}$: 200 $\mu m^2$). *Bottom row:* Corresponding cross-sectional views of each fabrication stage, as indicated by the vertical arrows, showing individual layer thicknesses and stack sequences.

## 2. Results and Discussion

### 2.1. bR SBL-Based Crosswire Device

We fabricated a protein-based, simple two-terminal crosswire device for the ETp study, as shown in Scheme 1. The detailed device fabrication procedure is described in the Experimental Section. A 4 $\mu$m wide bottom Au electrode (see Scheme 1) was patterned on a silicon wafer with the composition of ($Si/SiO_2/Cr/Au$) (see Section 4.1.1). Each bottom line-electrode connects between two contact pads (Figure 1A, Scheme 1B), and nine such isolated devices were integrated in each chip with optimized inter-electrode spacing. The self-assembled bR-SBL (Figure 2A) was deposited on a bottom electrode, functionalized with a cysteamine monolayer as a linker (details in Section 4.1.2; see also Figure 1B). Finally, ~50 $\mu$m wide top electrode (eC/Au) was evaporated across the linker functionalized protein modified bottom electrode. The overlap of the crosswires defines the geometric protein junction area ($A_{geo}$), which we estimate at ~$2\times10^2$ $\mu m^2$. For reference, one device is intentionally shorted on each chip.

In the top contact, the 10 nm carbon interleaved layer serves as a conductive barrier, and most importantly, it is a protective layer (Figure 1B). The evaporated carbon (eC) atoms/clusters are very reactive and are expected to react with the surface-exposed atoms of the protein and remain there, rather than penetrating further into it.[17] The deposited eC layer is effectively non-destructive, as reported earlier, where ultra-thin (< 5-8 nm) conjugated organic molecules retained their structural integrity based on spectroscopic evidence.[19,20] Indeed, the 10 nm eC layer was conformally coated and formed a dense layer without any pinhole, with rms roughness ~0.5 nm, which does not allow the penetration of the outermost top Au layer (see SI Figure S1). The 25 nm Au overlayer serves a dual purpose: (1) it prevents oxidation of the underlying carbon layer, preserving its conductivity, and (2) it provides the necessary conductance for electrical biasing and current transport. To minimize carbon oxidation and avoid insulating layer formation, the Au layer was deposited immediately after the carbon layer without breaking the vacuum (see Section 4.1.3). The composition of the evaporated carbon layer plays a crucial role in determining the device's performance; the higher the fraction of $sp^2$-hybridized carbon,[18] the higher the electrical conductivity.[17] The chemical composition was characterized using X-ray photoelectron spectroscopy, with additional details provided in the Supporting Information (see Section S1).

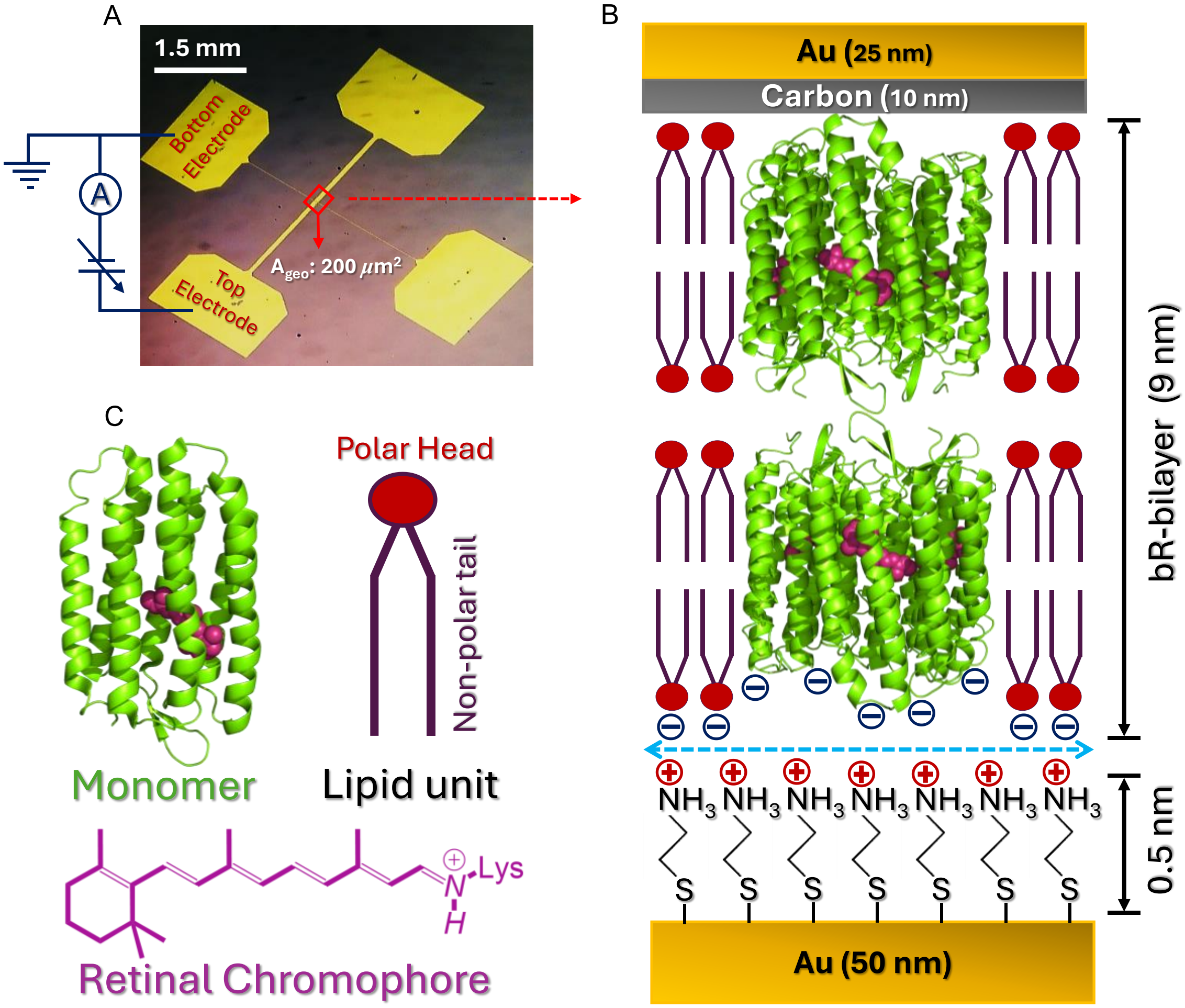

**Figure 1.** Schematic representation of Au/bR/eC/Au crosswire junction devices. (A) Optical microscopy image of a representative device with an experimental biasing scheme. The junction area ($A_{geo}$) is highlighted by a red square. (B) Schematic cross-section, showing the top and bottom electrodes, linker, and bR bilayer. The linker-protein electrostatic interactions at the interface are indicated by the horizontal blue dashed arrow. The bR trimer shows the 1BRR PDB model. (C) Structural components of bR, including the monomeric unit with its retinal protonated *Schiff base* chromophore (in purple), embedded in the peptide *α-helices*, and a schematic of one of the lipid molecules that surrounds the protein as one (vertical) bilayer structure, as shown in (B). (B) and (C) are adapted from ref.[9] and modified.

*2.1.1. Orientation and Selection of bR in Device Configuration*

bR is a remarkably stable trans-membrane protein known for its exceptional thermal and structural robustness. It maintains its native conformation over a wide temperature range, with a denaturation temperature exceeding 90 °C in solution, and notably, retains structural integrity even in the solid state up to approximately 140 °C.[28] This extraordinary thermal stability motivates the design and fabrication of evaporated top-contact protein junctions using bR as the active molecular component. In its native environment, bR exists as a trimer embedded within the purple bacterial membrane lipid matrix.[29] Upon treatment with octylthioglucoside (OTG), it forms doughnut-shaped vesicles with a narrow size distribution.[25] These vesicles are retained after OTG removal through dialysis, yielding a stable bR suspension in phosphate buffer (PB, pH 6.4).[9,25] Upon immobilization on the substrate and subsequent drying, these vesicles collapse into planar bilayer structures (see SI Figure S2).[9] This process yields high-quality, oriented, and uniform protein bilayers (i.e., double lipid bilayers) across large substrate areas.[9] Ellipsometric thickness measurements (~9 nm) and atomic force microscopy (AFM) characterization confirm the presence of the bR-supported lipid (bR-SBL) in a bilayer configuration.

The fabricated crosswire device contains a bR-SBL as illustrated in Figure 1. The bilayer is composed of ~70% protein, with the remaining ~30% native lipid junctions, both in terms of volume and (contact) areas.[25] Since native lipids are mainly saturated hydrocarbon chains, the lipid-only regions (~9 nm thick) will be highly resistive compared to the protein part, as expressed in their respective length decay parameter values, of $> 7$ nm$^{-1}$ [30] and $< 1$ nm$^{-1}$.[16,31]

## 2.2. Protein Layer Characterization

The stacked-layer device architecture inherently limits direct characterization of the protein layer (see Figure 1A). Therefore, the structural integrity and quality of the protein film were evaluated before device fabrication on the bR-SBL deposited on a linker-coated flat Au substrate, prepared identically to the device's bottom electrode (see Section 2.1). The detailed protein layer preparation protocol is given in the Experimental Section 4.1.2. The thickness of the deposited protein layer, determined by fitting ellipsometry results, as described in SI Section S2.1, was 9 ± 0.5 nm, in excellent agreement with earlier results.[9,16,32] AFM revealed uniform and dense protein coverage (>95%) across the substrate, with a surface roughness of 2.2 ± 0.1 nm and Z-height values consistent with typical bR-SBL films (see Figure 2A).[9,13,16,32] Detailed AFM techniques are described in SI Section S2.2.

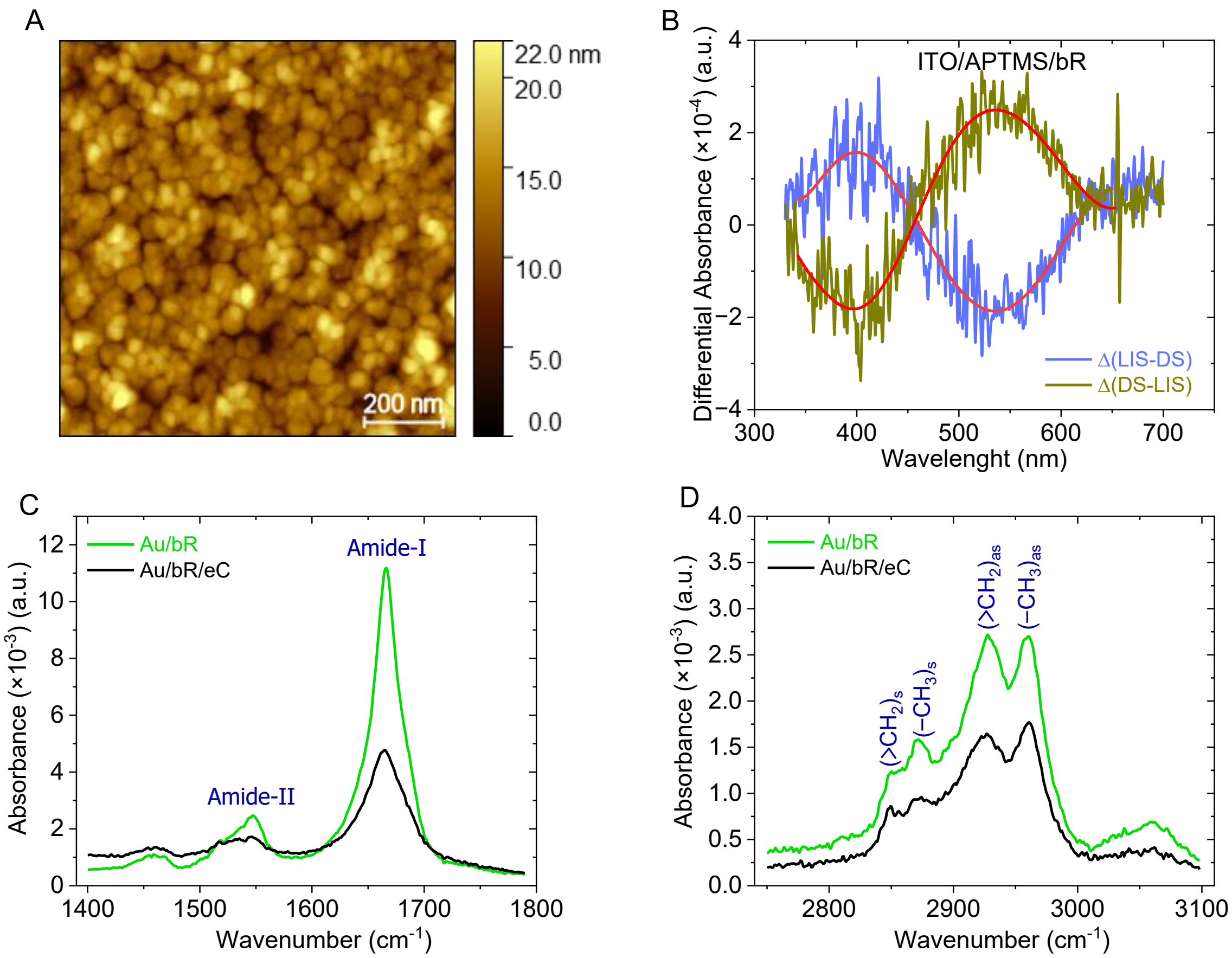

**Figure 2.** Characterization of bR protein layers. (A) Tapping-mode AFM topographic image of a single bR layer on a cysteamine-coated Au substrate. (B) Optical absorption *difference* spectra, showing the photocycle of a bR layer (as the one used in the electrical transport experiments), on an optically transparent ITO substrate, bonded via APTMS linker. The observed photocycle is shown as the combined difference (Δ) spectra between light-induced (LIS) and dark (DS) spectra,[13] where the red solid line shows the fitted data. (C, D): Representative PM-IRRAS spectra of a bR monolayer on cysteamine linker-coated Au, with and without a top coating with a transparent eC (10 nm)/Au (15 nm) layer, mentioned with the appropriate figure legends. (C) Amide I (~1666 cm$^{-1}$) and amide II (~1547 cm$^{-1}$) regions within 1400–1800 cm$^{-1}$. (D) Higher-wavenumber region (~2900 cm$^{-1}$), showing distinct C–H stretching vibrations.

*2.2.1. Characteristic Photocycle of bR in the Solid-State Configuration*

An important question is whether the protein, as a solid-state, i.e., dry film, still functions as in solution, despite the absence of its native buffer environment. In earlier work, we demonstrated that bR exhibits nearly identical UV–Vis absorption signatures in both solution and solid-state thin-film configurations.[33] Because we have a photo-active protein (with a long lifetime of a key intermediate), and this photoactivity is known from solution studies (since the native absorption spectrum remains unchanged) to require intact conformation,[26,34–36] we can use bR's light-driven structural isomerization, the *photocycle*, to check its conformational and functional integrity in the dry film, as discussed in detail earlier.[13,26,33]

Photocycle measurements were performed on a bR-SBL deposited on an optically transparent indium tin oxide (ITO) substrate. The bR-SBL was formed via electrostatic interactions using 3-aminopropyl trimethoxysilane (APTMS) as a linker (see Experimental Section 4.1.2). The detailed experimental procedure for photocycle measurements is described in ref.[13]. As shown in Figure 2B, the differential absorbance spectra display upon illumination a characteristic increase at ~410 nm and a corresponding decrease at ~570 nm, which thermally reverses in the dark. The formation of the intermediate M-state, characterized by the ~410 nm absorption peak under illumination (>500 nm wavelength),[26] and its thermal decay in darkness, confirms the conformational integrity of bR in a dry film (dry, except for its structural bound water).

Recent work by Bera *et al.*[13] showed that the *photocycle* of a solid-state bR film remains intact *even after evaporating a thin transparent top Pd electrode* in the glass/APTMS/bR(trilayer)/Pd configuration. There, a thicker (triple-bilayer or simply trilayer) bR film structure was used, which allows for the successful capturing of the transmitted light signals, as was shown by optical detection of the above-explained *photocycle* of bR. In the present work, a 3× thinner film of bR (~9 nm) was used with a thin, evaporated carbon (eC), top layer in the configuration of ITO/APTMS/bR/eC. Under these conditions, the change in optical signal, due to the photocycle, was below the detection threshold (data not shown). Still, this junction exhibited the photocycle effects on its ETp response, as discussed in detail in Section 2.4.2.

*2.2.2 Polarization Modulated Infrared Reflection Absorption Spectroscopy (PMIRRAS) Characterization*

The structural integrity of solid-state protein films can be evaluated by its infrared (IR) spectral features, particularly the positions of the amide-I (~1670 $cm^{-1}$) and amide-II (~1550 $cm^{-1}$) bands, along with characteristic C–H stretching vibrations (~2900 $cm^{-1}$).[13,37–39] PMIRRAS measurements were performed on bR-SBL films deposited on reflective gold (Au) substrates. The detailed PMIRRAS methodology, measurement conditions, and data analysis protocols are described in SI Section S2.3. The photocycle experiments discussed earlier in Section 2.2.1 show that protein functionality is preserved in the solid-state thin-film configuration. Based on this, the PMIRRAS spectrum of the bR-SBL film on Au (Au/Cys/bR-SBL) was considered the reference signal. Under identical conditions, PMIRRAS spectra were also recorded on bR-SBL films coated with a thin (~10 nm) semitransparent eC layer, deposited at the same rate used during device fabrication (see Experimental Section 4.1.3).

The resulting IR spectra (Figures 2C, 2D) show a clear overlap between the characteristic amide and C–H stretching bands of the carbon-coated and uncoated films, indicating that the protein's spectral features are well preserved. The reduced overall spectral intensity of the eC-coated sample is due to the strong IR absorption of the thin eC layer. The coated samples show relatively well-resolved amide bands, suggestive of surface-enhanced infrared absorption effects (SEIR),[40] consistent with our recent observations for protein films covered with semi-transparent metallic palladium layers.[13]

Notably, while the amide-I bands remained prominent, the amide-II features appeared weaker and approached the noise limit of the instrument. Both amide-I and amide-II regions (black spectra in Figures 2C, 2D) are known to comprise multiple sub-peaks (due to SEIR), as revealed by the spectral deconvolution that we have reported earlier.[38] The observed consistency in the amide-I and amide-II band positions, together with the preserved photocycle activity, are consistent with notion that the overall structural conformation of the protein also remains intact in the solid state, even after deposition of the top eC layer.

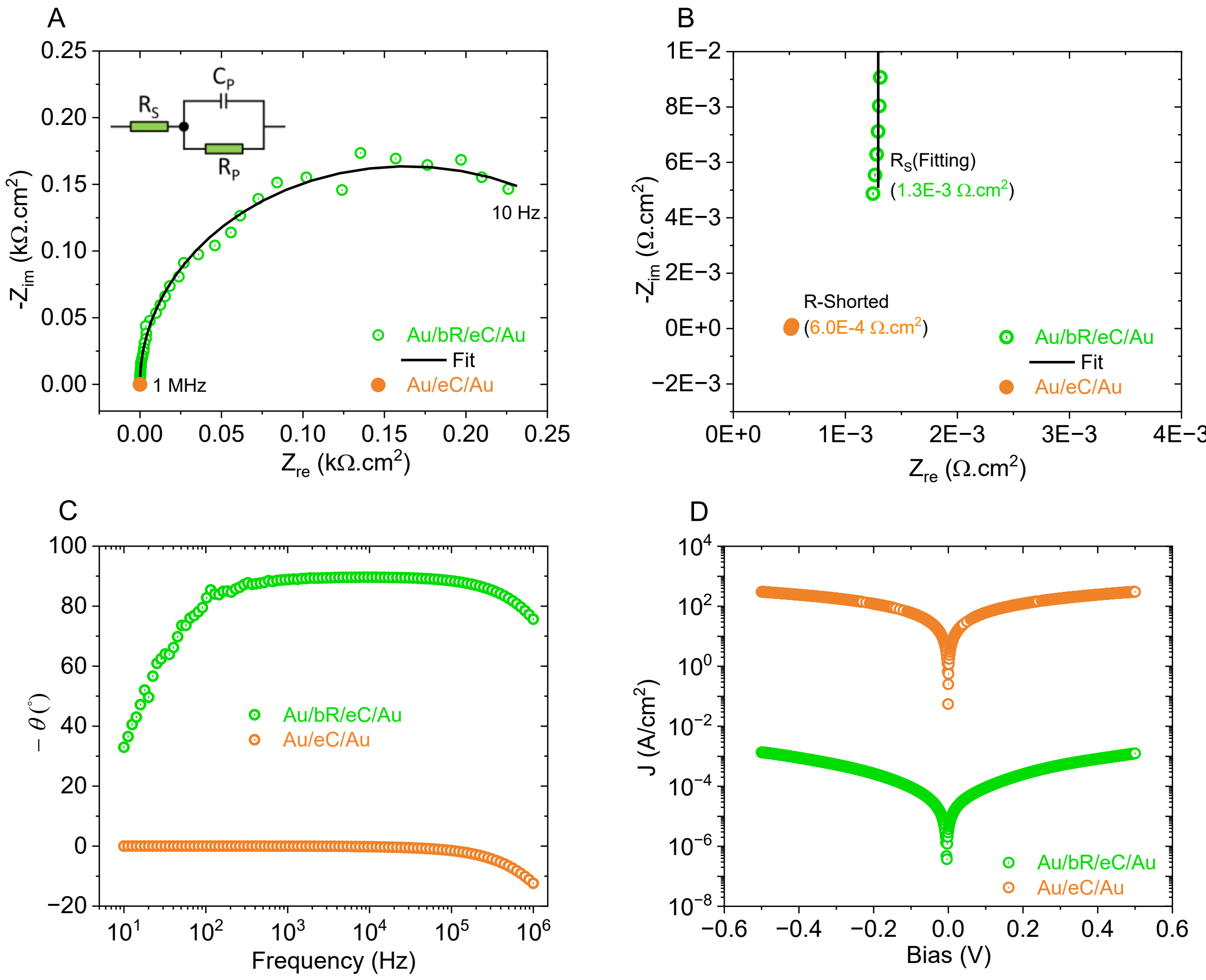

**Figure 3.** Impedance response of active (green) and shorted (orange) protein junctions. (A) Nyquist plots over the 10 Hz–1 MHz frequency range. The black solid line represents fitted data based on the equivalent circuit, shown in the top left inset of the figure. (B) Enlarged Nyquist plot highlighting the high-frequency region (~1 MHz). (C) Impedance phase (Bode) plots. (D) Corresponding J–V responses.

### 2.3. Transport-Active Protein Junctions: Insights from Impedance Spectroscopy

In this work, impedance spectroscopy was used to characterize the Au/bR/eC junctions. Each chip contained nine devices: five exhibited stable transport-active behavior, three were partially shorted (unstable), and one was intentionally shorted for control purposes. Details of impedance measurements, protocols, and data analysis are provided in SI Section S3.1. The focus here is on the transport-active junctions. The top eC electrode plays a crucial role in defining the junction properties. Evaporated carbon, consisting of reactive atomic or cluster species, readily interacts with organic surfaces but typically forms a surface-bound layer rather than penetrating into the organic matrix.[17] In our fabrication, a slow deposition rate ensured that the reactive carbon atoms form an effective (see Experimental Section 4.1.3), conformal top contact over the densely packed protein surface (>95% surface coverage, as confirmed by AFM), minimizing the risk of filament formation through the protein film.

The absence of a Warburg element in the low-frequency impedance (see Figure 3A) confirms that electrical transport through these dry solid-state protein junctions is purely electronic, with no ion diffusion contribution.[9,16] Protein-based electronic junctions, when sandwiched between two conductive electrodes, exhibit impedance characteristics analogous to a parallel resistor–capacitor (R–C) circuit, typically representing a single dielectric relaxation process.[9,16]

To elucidate the electrical behavior, impedance spectra were recorded for both transport-active (Au/bR/eC/Au) junctions and protein-layer-depleted fully shorted (Au/eC/Au) junctions. The comparison was carried out by analyzing:

(1) the resistance components obtained from Nyquist-plot fitting, and
(2) the impedance phase response.

*2.3.1. Nyquist Analysis and DC Correlation*

For the shorted junctions, the Nyquist plot is reduced to a single point, signifying a purely resistive behavior with no observable dielectric process (see SI in ref.[16]), confirming that the 10 nm eC layer is fully conductive and does not contribute any capacitive component. In contrast, the transport-active junctions exhibited a well-defined semicircular Nyquist profile (see Figure 3A), which fitted nicely ($\chi^2 < 10^{-2}$) to a simple R–C equivalent circuit[9,16] comprising a protein junction resistance ($R_P$) in parallel with a capacitance, connected in series with a contact resistance ($R_S$), as illustrated in the inset of Figure 3A.

The fitted parameters yielded $R_P \approx 3 \times 10^2$ Ω·cm$^2$, approximately $10^5 \times$ higher than what was measured with the shorted junction resistance without any intervening protein layer, $R_S \approx 1 \times 10^{-3}$ Ω·cm$^2$. This agrees with our experimental results, shown in Figure 3B, the zoomed-in Nyquist plot, where *R-shorted* is closely approaching to $R_S$ (fitted value); the deviation $R_S$/*R-shorted* ≈ 2 can be an experimental error consistence with our recent report.[16] This $R_S$ value is consistent with the intrinsic resistance of the contact leads and external circuitry, in agreement with other reports.[16] The corresponding DC current–voltage (*J–V*) characteristics of the same devices show more than $10^5$ × current differences between the transport-active and shorted junctions (see Figure 3D), confirming the strong correlation between AC and DC measurements and validating the electrical integrity of the bR-based devices.[16]

*2.3.2. Bode-Phase Response*

The impedance phase response (Bode analysis) offers an additional diagnostic for evaluating junction integrity. A high-frequency phase approaching 90° corresponds to ideal dielectric behavior, and a phase near 0° signifies a shorted or metallic (exclusively resistive component) response.[13,16] Thus, the phase responses clearly distinguish between transport-active and shorted junctions (see Figure 3C):

- *Shorted* Au/**eC**/Au junctions maintain a near-zero phase across the full frequency range, consistent with metallic conduction.
- *Transport-active* Au/**bR**/eC/Au junctions display a high-frequency phase approaching 90° above 200 Hz, reflecting strong capacitive behavior and confirming the formation of a high-quality dielectric barrier (protein layer) between the electrodes.

Thus, the impedance results clearly demonstrate that the ∼9 nm bR-SBL layer acts as an effective resistive–capacitive barrier,[9,13,16] suppressing direct electrical shorting and supporting electron transport exclusively through the protein matrix. Additional analysis provided in the *Supplementary Information* Section S3.2 compares the calculated and experimental current densities, further ruling out filamentary conduction and confirming the clean transport-active nature of the bR-SBL junctions.

## 2.4. Electron Transport in the Dark and under Illumination at Room Temperature

The direct current (DC)-based ETp measurements across the Au/bR/eC protein junctions were performed both in the dark and under illumination (with a laser of specific wavelengths) to investigate photo-induced electron transport behavior (refer to SI Section 4.1 for measurement protocol details).

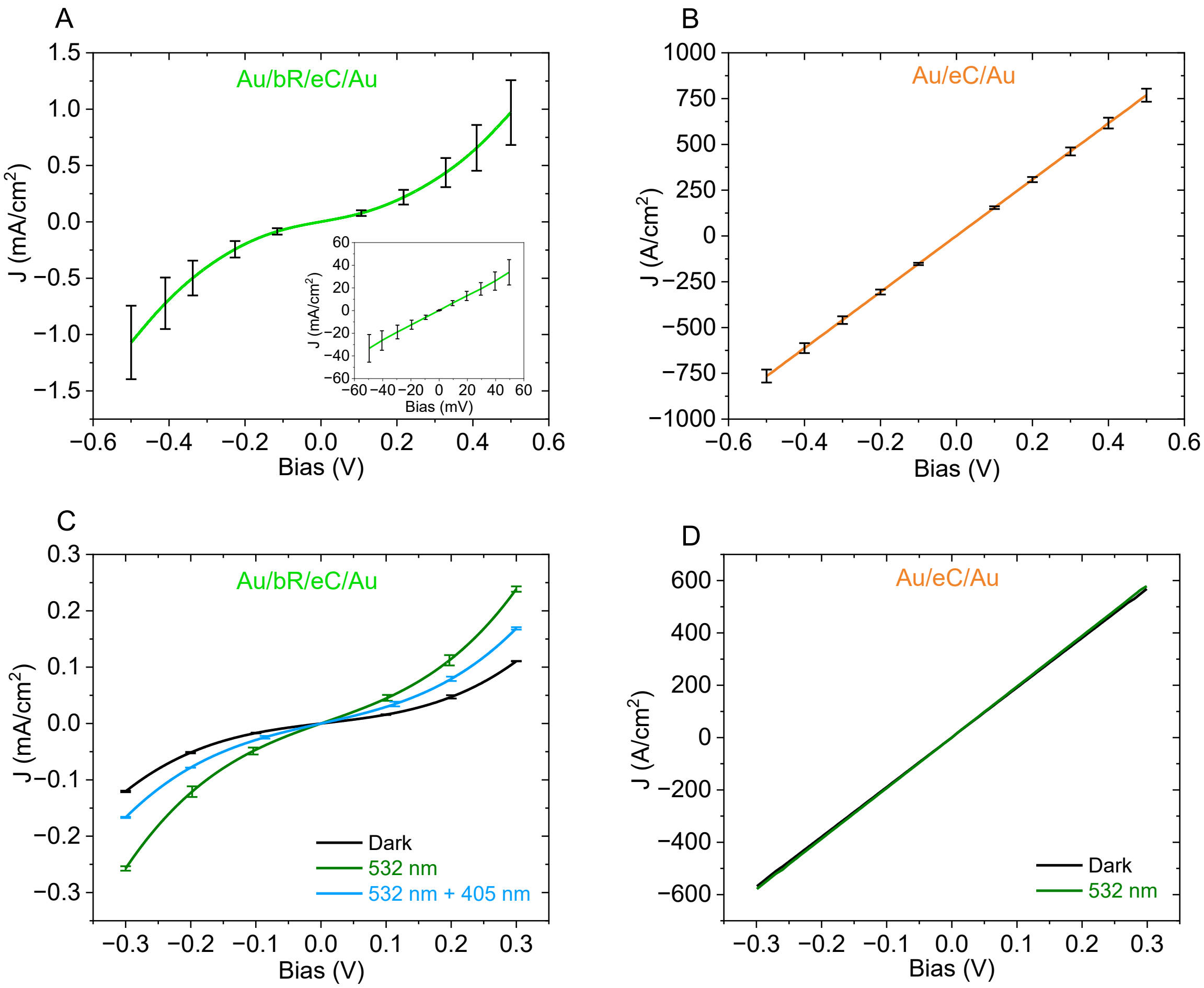

**Figure 4.** Current–voltage responses of the Au - eC/Au junctions with and without an interleaved protein layer under laser illumination and in the dark. (A) Averaged current density *(J)* –*V* response from 10 transport active protein junctions; error bars represent the SD around the mean (mean ± SD). *Inset*: linear *J*–*V* response of these junctions, over a narrow bias window (±50 mV), showing ohmic behavior. (B) Averaged *J*–*V* response of shorted junctions, i.e., without protein, from 5 devices; error bars represent mean ± SD. (C) Representative *J*–*V* response of protein junctions in the dark, under 532 nm laser, and under combined 532 nm and 405 nm laser illumination, shown with the color coding and legends. Error bars indicate ranges of values over two consecutive sweeps. (D) Control experiment on a shorted, i.e., protein-free device under light illumination, showing no significant current variation.

*2.4.1. Electron Transport without Illumination*

Figure 4A presents the averaged *J–V* characteristics of ten transport-active junctions in three independent chips. The current values fall within a narrow distribution (mean ± SD), confirming good device reproducibility. The *J–V* curves are nearly symmetric with respect to bias polarity, indicating uniform electrode–protein coupling. At low bias (±50 mV) (see inset of Figure 4A), the response approaches a linear, ohmic one, while nonlinearity emerges beyond ±0.1 V (see Figures 4A and 4C). In contrast, protein-free (shorted) junctions display a fully linear, ohmic *J–V* response across the entire voltage range examined (see Figures 4B, 4D, and S3), confirming that current flow in protein devices arises primarily from the bR layer rather than from the linker, electrode, or the interfaces. Control experiments using a ~0.5 nm cysteamine linker alone yielded 100% shorted junctions, with their *J–V* characteristics overlapping those of the bare electrodes (see SI Figure S3), further ruling out linker-related contributions. A similar observation was recorded in our recent work.[16]

The transport-active junctions retained their electrical characteristics over more than three months (see SI Figure S4), demonstrating long-term stability compared to previously reported micropore device (MpD) configurations, where most junctions degraded within a few weeks.[13] The improved stability is attributed to the better compatibility of the eC than the Pd top electrodes for protein devices.

*2.4.2. Photo-Induced Effect on Electron Transport*

To study photo-effects on the electron transport (PE-ETp), we fabricated optically semi-transparent Au/bR/eC/Au junctions composed with a thinner top Au layer (15 nm) to improve light penetration (see Experimental Section 4.1.3). The devices were measured sequentially in the dark, under illumination with a green laser (532 nm), then with combined green and blue laser beams (532 nm + 405 nm), and finally again in the dark. Upon green light illumination, the junction current doubled, i.e., $J_{green}/J_{dark} \approx 2$ (see Figure 4C). Subsequent exposure to dual-wavelength illumination reduced the current. Once the illumination is terminated, the current has thermally reverted to its original dark-state level. Control experiments on protein-free shorted junctions showed no measurable changes (see Figure 4D) in current, i.e., $J_{green}/J_{dark} \approx 1$, confirming that the photo-response originates from the protein layer.

The observed photo-induced current enhancement is consistent with the light-driven conformational changes and/or charge redistribution in bR associated with the relatively long-lived M-intermediate (due to illumination of green laser), which opens a more facile transport pathway across the protein film. The lifetime of the M intermediate in the solid state is much longer than in solution,[14,26] due to the absence of rapid thermal decay by the surrounding accessible water molecules. Thus, in solid-state, M-state accumulates to a high extent and promotes the ETp rate under illumination. Furthermore, the illumination of a 405 nm laser partially depopulates the M-state, restoring the protein to its initial stage and consequently reducing the current. Notably, the photo-response is not attributed to direct optical excitation of protein's electronic states, as the energy levels of bR are too far from the Fermi levels of the electrodes to be within the experimental bias window.[9] Thus, the photo-enhanced ETp can be attributed to M intermediate formation and accompanied conformation modulation[41] or charge[42] redistribution in the protein matrix, rather than electronic excitation. This situation is reminiscent of the effect of biotin binding to streptavidin,[38] which induces a structural rearrangement that modulates the current across the junction.

Overall, the reproducible light-induced enhancement in multiple independent devices ($J_{green}/J_{dark} \approx = 2.0–2.2$) provides compelling evidence that the bR-SBL retains its functional and structural integrity even in a solid-state evaporated-contact junction, establishing the viability of bR as a robust, light-responsive molecular component in bioelectronic devices.

### 2.5. Temperature-Dependent Transport and its Mechanism

The Au/bR/eC junctions exhibited remarkable stability and consistent transport activity over a broad temperature range (300 – 8 K), providing strong evidence of the robustness of the protein-based junctions. Details of the temperature-dependent ETp measurement protocol and setup are provided in the Supporting Information (Section S4.2). The ~9 nm bR-SBL films displayed nearly temperature-independent behavior. In the low temperature range, the currents are temperature-independent (Figure 5A), leading to an Arrhenius plot (ln $J$ vs $1/T$) with an essentially flat slope from ~ 100K down (Figure 5B).

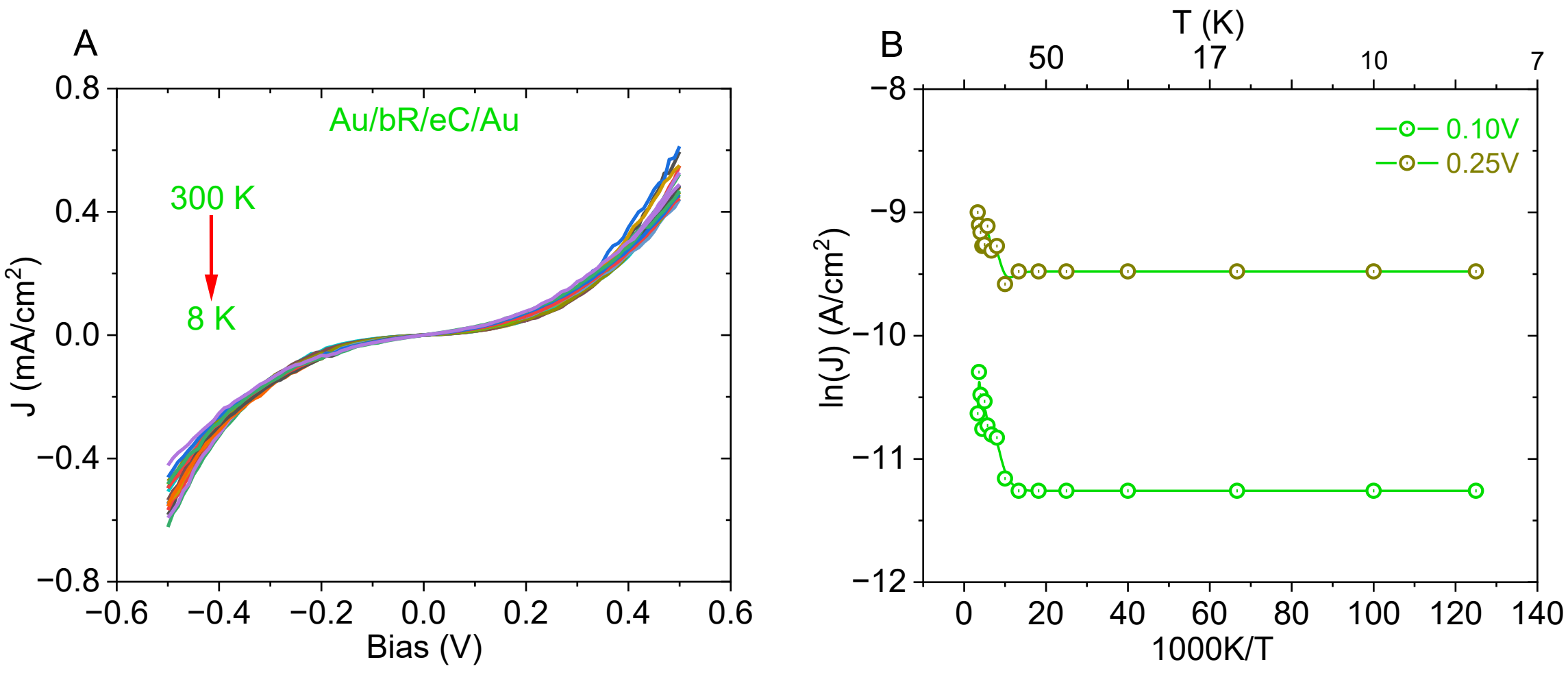

**Figure 5.** (A) Representative temperature-dependent *J–V* response (linear scale) of Au/bR/eC junctions measured between 300 K and 8 K in a setup with liquid helium cooling. (B) Corresponding *Arrhenius plots* under two different applied bias conditions, as indicated in the top right legends. The junction current decreases at low temperatures by a factor of less than 2 ($J_{300K}/J_{8K} < 2$).

In the high-temperature region, the estimated activation energies were below 10 meV across the entire experimental voltage range (±0.5 V), which amounts to $k_BT$ (8–26 meV). In the low-temperature regime (< 100 K), there is no transport activation energy within our experimental resolution (see Figure 5B). These results demonstrate that the ~9 nm bR-SBL junctions exhibit a thermally non-activated electron transport process, effectively ruling out any hopping-type conduction mechanism. Similar low-activation transport behavior has been reported previously for protein-based junctions with varying device architectures and different proteins, both by our group[9,37,43] and others[44]. Given the experimental junction length (~ 9 nm), conventional coherent tunneling is quantum-mechanically extremely improbable. Furthermore, the observed temperature-independent ETp responses are difficult to reconcile with nuclear tunneling models. Despite this fundamental challenge to establishing a complete microscopic picture, the demonstrated solid-state, temperature-resilient protein junctions represent a promising platform for the development of future bioelectronic devices.

### 2.6. Advantages and Limitations of the Au/bR/eC Junction

The problem of considering conventional electron transport (ETp) mechanisms, namely hopping and tunneling, to these and other protein-based junctions may be helped by results of even lower temperature measurements of contact-resistance - free junctions, at temperatures where protein vibrations will be affected. This is where protein junction configurations like the one explored here, robust and transport-active, are essential for such lower temperature investigations.

The present study demonstrates that the Au/bR/eC device exhibits high stability over both extended timescales (over 3 months) and a broad temperature range (down to 8K from RT). This robustness enables a direct evaluation of intrinsic protein transport characteristics. The use of a carbon top electrode provides a compatible organic–carbon interface that minimizes contact resistance,[16,27] thereby improving measurement reliability. Based on the *J-V* response influenced by the photo-induced structural isomerization of bR even with the evaporated eC top contact, bR-SBL retains functional activity inside the junction, (see Section 2.4.2). Importantly, the evaporated carbon electrode allows for the fabrication of non-shorted, transport-active junctions, achieving yields exceeding 40%. This performance is the real advantage over the recently reported metal–protein–device (MpD) architectures based on permanent metallic contacts, where only a thicker protein multilayer yields non-shorted devices. In addition to the ~9 nm bR assemblies, we have also integrated other well-characterized protein monolayers, including azurin (~2.4 nm) and photosystem I (~5 nm), within the Au-eC device structure, achieving similarly high yields of transport-active junctions (data not shown, unpublished).

Despite these advantages, the Au/protein/eC configuration presents certain limitations. The relatively low conductivity of the evaporated carbon (eC) electrode compared to metallic contacts, such as those used in MpD systems, constrains its suitability for studying multilayer protein assemblies and length-dependent transport phenomena. Moreover, although the top electrode is evaporated, precise control of the effective junction area remains challenging. The geometric over-lap area between the top and bottom line electrodes ($A_{geo} \approx 200\ \mu m^2$) leads to an estimated dielectric constant roughly 50 times greater than the reported intrinsic value (~5) for bR,[9] as inferred from impedance-derived capacitance ($C_P$) measurements in earlier studies. This discrepancy would be removed if the active area were 50 × smaller, i.e., 4 $\mu m^2$, highlighting the well-known problem of biomolecular ensemble junctions, the difference between the electrically active junction area and the geometric area.

## 3. Conclusions

In this work, we have established a stable, temperature-resilient, and photo-responsive protein-based electronic junction using bacteriorhodopsin (bR) integrated within an Au/bR/eC architecture, which provides a versatile and reproducible route toward a functional molecular component (protein) entrapped in bioelectronic devices, highlighting the robustness of bR that bridges the molecular and electronic domains. The top carbon contact forms an electronically benign and mechanically compatible interface, enabling non-shorted, transport-active junctions even with single protein layers. The temperature-independent ETp behavior, combined with reversible photo-induced modulation of current, demonstrates a non-thermally activated transport mechanism mediated by conformational dynamics within the protein matrix. The demonstrated system represents a major step toward realizing solid-state protein-based electronics that maintain structural integrity and transport activity over cryogenic and ambient conditions alike. Nonetheless, challenges remain in precisely defining the active junction area and improving electrode conductivity for multilayer and length-dependent studies.

## 4. Experimental Section

### 4.1 Crosswire Au/bR/eC Device Fabrication

#### 4.1.1. *Bottom Electrode Fabrication*

Bottom electrodes were fabricated on a Si(100) substrate coated with a 300 nm thermally grown $SiO_2$ insulating layer. A standard photolithography process[13] defined 4 $\mu$m wide Au lines and four large contact pads (Scheme 1B). Metal deposition was performed using e-beam physical vapor deposition (PVD, *Angstrom Evaporator*) at a base pressure of < $1 \times 10^{-7}$ mbar. A 3 nm Cr adhesion layer and a 50 nm Au layer were deposited sequentially at rates of 0.2 Å $s^{-1}$ and 0.5 Å $s^{-}$ [1], respectively, with a 300 mm source–substrate distance. In a similar deposition rate and Cr/Au layer composition, a flat Au substrate was prepared on the same silicon wafer mentioned above. Liftoff in acetone defined the final electrode pattern.[13] Each 1.5 × 1.5 $cm^2$ chip contained nine electrically isolated device arrays.

4.1.2. *bR Self-Assembled Layer (bR–SBL) Preparation*

Prior to molecular assembly, the patterned substrates were cleaned by sequential boiling in acetone and isopropanol (IPA), rinsed in Milli-Q water, and dried under $N_2$. Surface functionalization was achieved by immersing the Au-patterned chip in a 10 mg $mL^{-1}$ aqueous solution of cysteamine hydrochloride (Sigma) for 12 h at room temperature, followed by thorough Milli-Q water rinsing and $N_2$ drying. The thiol terminus of cysteamine forms a covalent bond with Au, while the terminal $–NH_3^+$ group provides a positively charged surface (pH 6.4–7).

A 4 $\mu$M solution of bacteriorhodopsin (bR) in 10 mM phosphate buffer (pH 6.4) containing $(NH_4)_2SO_4$ was drop-cast selectively on eight device regions, leaving one as an unmodified control. The samples were kept in a humidity-controlled chamber for 12 h to prevent dehydration and ensure electrostatic adsorption of the negatively charged bR onto the cysteamine layer. After incubation, the chips were rinsed with Milli-Q water and dried under $N_2$.

For characterization, bR–SBL films were prepared under identical conditions on flat Si/Cr (3 nm)/Au (50 nm) substrates for PMIRRAS, ellipsometry, and AFM measurements. For photocycle experiments, bR–SBL was assembled on ITO/quartz substrates cleaned by sequential ultrasonication (10 min each in acetone, ethanol, and water) and $N_2$ drying. The ITO surface was silanized using 3-aminopropyl trimethoxysilane (APTMS, MeOH : $H_2O$ : APTMS = 3000 : 100 : 50 µL) for 6 h, yielding an amine-terminated surface analogous to the cysteamine-functionalized Au. The bR layer was then formed by 12 h incubation in 4 $\mu$M bR solution, followed by rinsing and drying.

4.1.3 *Top eC/Au Electrode Deposition*

Protein-coated substrates were mounted in the evaporator with a cross-aligned shadow mask relative to the bottom electrodes (Scheme 1D). After overnight pumping to achieve $< 1 \times 10^{-7}$ mbar, the 50 $\mu$m wide top electrode was deposited with a 10 nm evaporated carbon (eC) film, maintaining a deposition rate of 0.1 Å $s^{-1}$ using a pure graphite source (*SPI Supplies*). Without breaking the vacuum, a 25 nm Au layer was subsequently deposited at 0.5 Å $s^{-1}$. During deposition, the substrate was cooled to 10 °C and positioned 500 mm above the source.

For devices intended for photo-enhanced electron transport (PE-ETp) measurements, the Au capping thickness was reduced to 15 nm (eC = 10 nm) to enhance optical transparency. For PMIRRAS measurements, only the 10 nm eC top contact was deposited to avoid optical interference from Au.

**Supporting Information**

The Supporting Information (SI) provides additional experimental details and characterization data for this study. Section S1 presents X-ray photoelectron spectroscopy (XPS) analysis to verify the elemental composition and chemical states of the sample surfaces. Section S2 focuses on surface characterization, including thickness determination by ellipsometry (S2.1), surface morphology analysis by AFM (S2.2), and PM-IRRAS measurements of protein thin films (S2.3). Section S3 describes the impedance measurement setup (S3.1) and provides insights into possible filamentary pathways and their (im)probability in Au/bR/eC junctions, based on our results for the impedance phase (S3.2). Section S4 details the room temperature and temperature-dependent DC electrical measurements, with subsections on DC *J-V* characteristics (S4.1) and temperature-dependent studies (S4.2). Finally, all supporting figures and tables are compiled in the SI Figures and Tables sections (S8–S11).

**Author Contributions**

S.K.S., D.C., M.S., and I.P. conceptualized the project. S.K.S. guided device fabrication and was responsible for its optimization. All transport measurements, including those under cryogenic and laser-illuminated conditions, were carried out by S.K.S. and S.B. The PM-IRRAS and AFM-based characterizations were performed by S.B with guidance from M.S. T.B. was responsible for the XPS study and data analysis. S.K.S. designed the DC transport experiments, while all impedance measurements were conducted by S.B. Both S.K.S. and S.B. contributed to data extraction, analysis, and presentation. S.B., M.S, and D.C. drafted the manuscript with input from all authors.

**Acknowledgements**

We gratefully acknowledge the nanofabrication unit at the Chemical Research Support Department

(WIS), particularly Assaf Hazzan, Sharon Garusi, and Leonid Tunik, for their guidance and assistance in device fabrication. We thank Dr. Omer Yaffe for providing access to impedance measurement facilities. We thank Drs. R. McCreery (University of Alberta) and J. A. Fereiro (IISER TVM) for insightful discussions. SKS acknowledges fellowship support from the Weizmann Institute, WIS, and its Graduate School. This work was supported by grants from the Tom and Mary Beck Center for Advanced and Intelligent Materials and the Kimmelman Center for Biomolecular Structure and Assembly, both at the Weizmann Institute of Science, where MS holds the Katzir-Makineni Chair in Chemistry.

## References

[1] I. Willner, E. Katz, *Bioelectronics: From Theory to Applications*, John Wiley & Sons, **2006**.
[2] A. Zhang, C. M. Lieber, *Chem. Rev.* **2016**, *116*, 215.
[3] B. J. Kim, Y. Ko, J. H. Cho, J. Cho, *Small* **2013**, *9*, 3784.
[4] X. Qiu, R. C. Chiechi, *Nat Commun* **2022**, *13*, 2312.
[5] M. Berggren, A. Richter-Dahlfors, *Advanced Materials* **2007**, *19*, 3201.
[6] B. Gürbüz, F. Ciftci, *Chemical Engineering Journal* **2024**, *489*, 151230.
[7] A. S. Shekhawat, B. Sahu, A. Diwan, A. Chaudhary, A. M. Shrivastav, T. Srivastava, R. Kumar, S. K. Saxena, *ACS Sens.* **2024**, *9*, 5025.
[8] P. Li, S. Bera, S. Kumar-Saxena, I. Pecht, M. Sheves, D. Cahen, Y. Selzer, *Proceedings of the National Academy of Sciences* **2024**, *121*, e2405156121.
[9] S. Bera, J. A. Fereiro, S. K. Saxena, D. Chryssikos, K. Majhi, T. Bendikov, L. Sepunaru, D. Ehre, M. Tornow, I. Pecht, A. Vilan, M. Sheves, D. Cahen, *J. Am. Chem. Soc.* **2023**, *145*, 24820.
[10] T. Jiang, B.-F. Zeng, B. Zhang, L. Tang, *Chemical Society Reviews* **2023**, *52*, 5968.
[11] C. Shipps, H. R. Kelly, P. J. Dahl, S. M. Yi, D. Vu, D. Boyer, C. Glynn, M. R. Sawaya, D. Eisenberg, V. S. Batista, N. S. Malvankar, *Proceedings of the National Academy of Sciences* **2021**, *118*, e2014139118.
[12] H. B. Gray, J. R. Winkler, *Chemical Science* **2021**, *12*, 13988.
[13] S. Bera, E. Mishuk, P. Li, S. Das, S. Keshet, S. Garusi, L. Tunik, E. Edri, Y. Selzer, I. Pecht, A. Vilan, M. Sheves, D. Cahen, *Small* ***2025***, e06560. https://doi.org/10.1002/smll.202506560
[14] C. D. Bostick, S. Mukhopadhyay, I. Pecht, M. Sheves, D. Cahen, D. Lederman, *Rep. Prog. Phys.* **2018**, *81*, 026601.
[15] D. Cahen, I. Pecht, M. Sheves, *J. Phys. Chem. Lett.* **2021**, *12*, 11598.
[16] S. Bera, A. Vilan, S. Das, I. Pecht, D. Ehre, M. Sheves, D. Cahen, *Advanced Materials* ***2025***, e07654.https://doi.org/10.1002/adma.202507654
[17] H. Yan, A. J. Bergren, R. L. McCreery, *J. Am. Chem. Soc.* **2011**, *133*, 19168.
[18] A. Morteza Najarian, B. Szeto, U. M. Tefashe, R. L. McCreery, *ACS Nano* **2016**, *10*, 8918.
[19] A. Morteza Najarian, A. Bayat, R. L. McCreery, *J. Am. Chem. Soc.* **2018**, *140*, 1900.
[20] S. K. Saxena, U. M. Tefashe, R. L. McCreery, *J. Am. Chem. Soc.* **2020**, *142*, 15420.
[21] A. S. Shekhawat, N. K. A. B, A. Diwan, D. Murugan, A. Chithravel, L. Daukiya, A. M. Shrivastav, T. Srivastava, S. K. Saxena, *Nanoscale* **2025**, *17*, 8363.

[22] A. S. Shekhawat, A. Diwan, T. Srivastava, R. Kumar, S. K. Saxena, R. L. McCreery, *J. Am. Chem. Soc.* **2025**, *147*, 27122.
[23] S. K. Saxena, U. M. Tefashe, M. Supur, R. L. McCreery, *ACS Sens.* **2021**, *6*, 513.
[24] S. K. Saxena, S. R. Smith, M. Supur, R. L. McCreery, *Advanced Optical Materials* **2019**, *7*, 1901053.
[25] N. D. Denkov, H. Yoshimura, T. Kouyama, J. Walz, K. Nagayama, *Biophysical Journal* **1998**, *74*, 1409.
[26] Y. D. Jin, N. Friedman, M. Sheves, D. Cahen, *Advanced Functional Materials* **2007**, *17*, 1417.
[27] A. Santos, U. M. Tefashe, R. L. McCreery, P. R. Bueno, *Physical Chemistry Chemical Physics* **2020**, *22*, 10828.
[28] Y. Shen, C. R. Safinya, K. S. Liang, A. F. Ruppert, K. J. Rothschild, *Nature* **1993**, *366*, 48.
[29] Y. Jin, N. Friedman, M. Sheves, T. He, D. Cahen, *Proceedings of the National Academy of Sciences* **2006**, *103*, 8601.
[30] R. Campos, R. Kataky, *J. Phys. Chem. B* **2012**, *116*, 3909.
[31] D. Chryssikos, J. A. Fereiro, J. Rojas, S. Bera, D. Tüzün, E. Kounoupioti, R. N. Pereira, C. Pfeiffer, A. Khoshouei, H. Dietz, M. Sheves, D. Cahen, M. Tornow, *Advanced Functional Materials* **2024**, *34*, 2408110.
[32] K. Garg, S. Raichlin, T. Bendikov, I. Pecht, M. Sheves, D. Cahen, *ACS Appl. Mater. Interfaces* **2018**, *10*, 41599.
[33] I. Ron, L. Sepunaru, S. Itzhakov, T. Belenkova, N. Friedman, I. Pecht, M. Sheves, D. Cahen, *J. Am. Chem. Soc.* **2010**, *132*, 4131.
[34] J. Wang, M. A. El-Sayed, *Biophysical Journal* **2000**, *78*, 2031.
[35] R. H. Lozier, R. A. Bogomolni, W. Stoeckenius, *Biophysical Journal* **1975**, *15*, 955.
[36] I. Rousso, N. Friedman, M. Sheves, M. Ottolenghi, *Biochemistry* **1995**, *34*, 12059.
[37] J. A. Fereiro, G. Porat, T. Bendikov, I. Pecht, M. Sheves, D. Cahen, *J. Am. Chem. Soc.* **2018**, *140*, 13317.
[38] S. Bera, S. Govinda, J. A. Fereiro, I. Pecht, M. Sheves, D. Cahen, *Langmuir* **2023**, *39*, 1394.
[39] J. G. Reji, M. Michael, A. Sivankutty, A. K. Rasheed, G. Krishna, M. Sheves, J. A. Fereiro, *Small* **2025**, *21*, 2502111.
[40] J. Kozuch, K. Ataka, J. Heberle, *Nat Rev Methods Primers* **2023**, *3*, 1.
[41] R. Orłowski, J. A. Clark, J. B. Derr, E. M. Espinoza, M. F. Mayther, O. Staszewska-Krajewska, J. R. Winkler, H. Jędrzejewska, A. Szumna, H. B. Gray, V. I. Vullev, D. T. Gryko, *Proceedings of the National Academy of Sciences* **2021**, *118*, e2026462118.
[42] C. Guo, Y. Gavrilov, S. Gupta, T. Bendikov, Y. Levy, A. Vilan, I. Pecht, M. Sheves, D. Cahen, *Physical Chemistry Chemical Physics* **2022**, *24*, 28878.
[43] K. Garg, M. Ghosh, T. Eliash, J. H. van Wonderen, J. N. Butt, L. Shi, X. Jiang, F. Zdenek, J. Blumberger, I. Pecht, M. Sheves, D. Cahen, *Chemical Science* **2018**, *9*, 7304.
[44] K. S. Kumar, R. R. Pasula, S. Lim, C. A. Nijhuis, *Advanced Materials* **2016**, *28*, 1824.

# Supporting Information

## Protein-Based Electrical Junctions with Robust Biocompatible Carbon Electrodes Exhibit Activation-less Charge Transport down to 10 K

Shailendra K. Saxena,[†,¥,€] Sudipta Bera,[†,¥] Tatyana Bendikov,[§] Israel Pecht,[‡] Mordechai Sheves,[†,*] and David Cahen[†,*]

[†]Department of Molecular Chemistry and Materials Science, Weizmann Institute of Science, Rehovot 7610001, Israel

[§]Department of Chemical Research Support, Weizmann Institute of Sciences, Rehovot 7610001, Israel

[‡]Department of Immunology and Regenerative Biology, Weizmann Institute of Science, Rehovot 7610001, Israel

[€]Department of Physics and Nanotechnology, College of Engineering and Technology, SRM Institute of Science and Technology, Kattankulathur, Chennai 603203, Tamil Nadu, India

[¥]The authors contributed equally to this work

[*]Corresponding Authors' email: *mudi.sheves@weizmann.ac.il*, *david.cahen@weizmann.ac.il*

**Supporting Information Index**

## S1. X-ray Photoelectron Spectroscopy (XPS) Analysis

XPS measurements were performed using a *Kratos AXIS ULTRA* system equipped with a monochromatic Al Kα X-ray source (hν = 1486.6 eV) operated at 75 W. Spectra were acquired with pass energies ranging from 20 to 80 eV. Curve fitting was carried out using linear or Shirley background subtraction and Gaussian–Lorentzian peak shapes.

In this study, we focused on the C1s core-level spectra of the eC surface. To establish a reference, the C1s signal was measured from both the surface, before and after eC deposition on the standard flat Au substrate (see **Experimental: Bottom Electrode Fabrication**). A weak carbon signal originating from surface contamination was observed on the bare Au substrate, whereas a significantly stronger signal appeared after eC deposition (**Figure S5**). To isolate the true contribution from the deposited eC, the pre-deposition spectrum was subtracted from the post-deposition spectrum (**Figure S5B**). The corrected C1s spectrum was deconvoluted into components corresponding to different carbon bonding environments.[1] The relative abundance of $sp^2$ carbon ($C(sp^2)$) is of particular importance, as it governs the electrical conductivity of the eC layer. Peak fitting revealed a high $sp^2$ carbon content (~62%), with an estimated $C(sp^2)/C(sp^3)$ ratio of ~3 in terms of atom percentage (see **Table S1**).[2] This strongly indicates that the eC film possesses a predominantly graphitic character and is highly conductive (**Table S1**).

Indeed, the eC layer deposited on top of the protein film is expected to be more conducive than the carbon layer analyzed in the XPS study. In the device, the carbon layer was deposited under controlled conditions and immediately covered with the protective Au top layer without breaking the vacuum ($<10^{-7}$ mbar), thereby preserving its high percentage $sp^2$ content. In contrast, for the XPS analysis, only a carbon layer was deposited directly on the Au substrate. Importantly, this sample was exposed to ambient conditions during transfer for XPS measurements setup, which likely led to some extent to partial oxidation and a significant reduction in the $C(sp^2)$ content.

## S2. Surface Characterization

Comprehensive surface characterization was performed to evaluate the quality and integrity of the deposited bR–SBL protein films, consistent with our previous studies. In this work, *ellipsometry* and *atomic force microscopy (AFM)* were employed to probe the film morphology and thickness, while *polarization modulation–infrared reflection–absorption spectroscopy (PM-IRRAS)* was utilized to confirm the characteristic chemical signatures of the protein structure.

### S2.1. Thickness Determination by Ellipsometry

Successive spectroscopic ellipsometry measurements (Ψ–Δ versus wavelength) were conducted on the flat Au substrate, the cysteamine-modified surface, and the subsequently formed bR–SBL layer over a wavelength range of 350–1000 nm at a fixed incidence angle of 70°, under ambient conditions (RH < 60%, $T$ = 293 ± 2 K). The optical response of the bare Au substrate was first modeled using a standard gold layer with variable $n$ and $k$ parameters. Subsequently, an independent *Cauchy model*[3,4] was applied in a stepwise manner to fit the cysteamine linker and bR–SBL layers using the built-in analysis software of the *Woollam M2000V* ellipsometer. Each fitting step provided the respective thickness of the individual layers (linker and protein film).[4]

### S2.2. Surface Morphology by AFM

Tapping-mode AFM imaging was performed to examine the surface topography and roughness of both the bR–SBL and the 10 nm evaporated carbon (eC) layer using a *Bruker Nanoscope V Multimode AFM*. Measurements were conducted under ambient conditions (RH < 40%, $T$ = 295 ± 5 K) with standard cantilevers (spring constant = 2 N $m^{-1}$, resonance frequency = 60–80 kHz). Image processing and quantitative roughness analysis were performed using *Gwyddion 2.64* software.

The AFM topography revealed a conformal, uniform, and continuous eC coating over the Au substrate, exhibiting a low *rms* roughness of 0.5 ± 0.05 nm (**Figure S1**). The morphology of the protein layer is presented in **Figure 2A** (main text).

### S2.3. PM-IRRAS Analysis of Protein Thin Films

To verify the structural integrity of the protein layer, PM-IRRAS measurements were carried out on bR–SBL films deposited on reflective Au substrates using a *Nicolet 6700* spectrometer equipped with a Photo-Elastic (PEM) module, grazing angle accessory, and a liquid nitrogen–cooled MCT detector. Each spectrum represents the average of 2000 scans acquired at 0.8 $cm^{-1}$ resolution with an 80° incident angle. Data acquisition and processing were performed using *Omnic 8.1* software.

Particular attention was given to the amide I and II band positions and the C–H stretching vibrations, whose relative intensities provide insights into the preservation of the protein's secondary structure. The comparative analysis confirms that the bR–SBL film retains its structural features even after the deposition of the top contact layer (see **Section 2.2.2**, main text).

## S3. Setup and Insights into Impedance

### S3.1. Impedance Measurement Setup

Impedance spectroscopy (IS) measurements for the Au/bR/eC junctions were conducted at room temperature (293 ± 2 K) using a *Zurich Instruments MFIA* impedance analyzer. The devices, mounted on a *LakeShore* probe station, were measured in a two-probe configuration under a vacuum of ~$10^{-2}$ mbar. An AC excitation of 10 mV was applied with zero DC bias, and the frequency was swept from 10 Hz to 1 MHz at a rate of 15 points per decade.

The resulting impedance spectra was analyzed in both Nyquist and Bode representations. Equivalent circuit modeling was performed using *ZView 4* (Scribner Associates), employing a parallel resistor–capacitor (*R–C*) network with an additional series resistance element.[4,5] The experimental data exhibited excellent agreement with the proposed equivalent circuit, yielding high-quality fits with $\chi^2 < 10^{-2}$.

### S3.2. Filamentary Pathways and Their Improbability in Au/bR/eC, Considering Impedance Phase

The electrical behavior of shorted protein junctions can be rationalized using a modified parallel *R–C* circuit (**Figure R6**), incorporating an additional resistance component ($R^*$) in parallel with the intrinsic protein resistance ($R_P$) and capacitance ($C_P$) (see **Section 2.3.1** in the main text). As $R^*$ represents a conductive filament pathway, it is expected to be several orders of magnitude smaller than $R_P$, though comparable to the series resistance ($R_S$).

Under a high-frequency AC field, defective junctions containing such filaments behave as simple resistive elements with a series combination of $R_S$ and $R^*$ (net-resistance = $R_S + R^*$; see **Figure S6B**). Since $(R_S + R^*) \ll R_P$, effectively rendering capacitive reactants and $R_P$ insignificant. Consequently, the impedance phase angle approaches zero across all frequencies, indicating a purely resistive response. Thus, the presence of conductive filaments (e.g., Au or C) through voids in the protein layer, characterized by $R^* \ll R_P$, can be identified through phase analysis. Experimentally, the observed phase response confirms that the Au/bR/eC junctions maintain complete dielectric separation between electrodes. (Further discussion is provided in **Section 2.3**, main text.)

To evaluate the plausibility of filamentary conduction, a simple resistance-based estimation was performed for a hypothetical carbon filament bridging across the ~9 nm protein junction. Assuming a minimal cross-sectional area of 1 $nm^2$ and a bulk carbon resistivity (considering upper limit) of $6 \times 10^{-4}$ Ω·m, the estimated filament resistance is ~5.4 MΩ, corresponding to a current of ~100 nA at 0.5 V, approximately two orders of magnitude higher than the experimentally observed junction current (~1 nA). To reconcile this discrepancy, the filament cross-section would need to shrink to an unrealistic ~1 $Å^2$, approaching the scale of a one-dimensional atomic chain. At such dimensions, quantum confinement effects dominate, and the expected current would be comparable to or even lower than that through the protein matrix itself.

Furthermore, the macromolecular packing of proteins in self-assembled bR–SBL films forms regular interstitial voids at the nanometer scale, implying that multiple filaments would form rather than one, leading to even higher total current than observed. Hence, both quantitative estimation and experimental evidence conclusively indicate that current transport in our protein junctions is not governed by filamentary conduction.

## S4. DC and Temperature Dependence

All electrical measurements, including DC and temperature-dependent studies, were performed on both bR–SBL and control (bare Au/eC) devices in a two-probe configuration inside a *LakeShore TTPX* probe station under high vacuum ($10^{-5}$ mbar).

### S4.1. DC Measurements

Room-temperature (RT) DC *J–V* characteristics were measured using a Keithley 6430 sub-femtoamp source meter. The bias voltage applied to the eC/Au top electrode varied from 0.1 V (2 mV step) to ±1.0 V (10 mV step), with the bottom Au electrode grounded. A complete voltage sweep cycle was executed, starting from 0 V → negative bias maximum → positive bias maximum → back to 0 V, using custom *LabVIEW*-based control software. The scan rate was 20 mV $s^{-1}$, and all measurements were conducted at 293 ± 2 K under ~$10^{-5}$ mbar vacuum.

### S4.2. Temperature-Dependent Measurements

Temperature-dependent current–voltage (*I–V*) measurements (±0.5 V) were carried out using liquid helium cooling in the temperature range 300 K → 8 K, with intermediate steps at 275, 250, 225, 200, 175, 150, 125, 100, 75, 55, 40, 25, 10, and 8 K. Temperature was precisely controlled using a *LakeShore 336* temperature controller equipped with a *DT421* silicon diode sensor attached near the sample stage, achieving an accuracy of ±5 K.

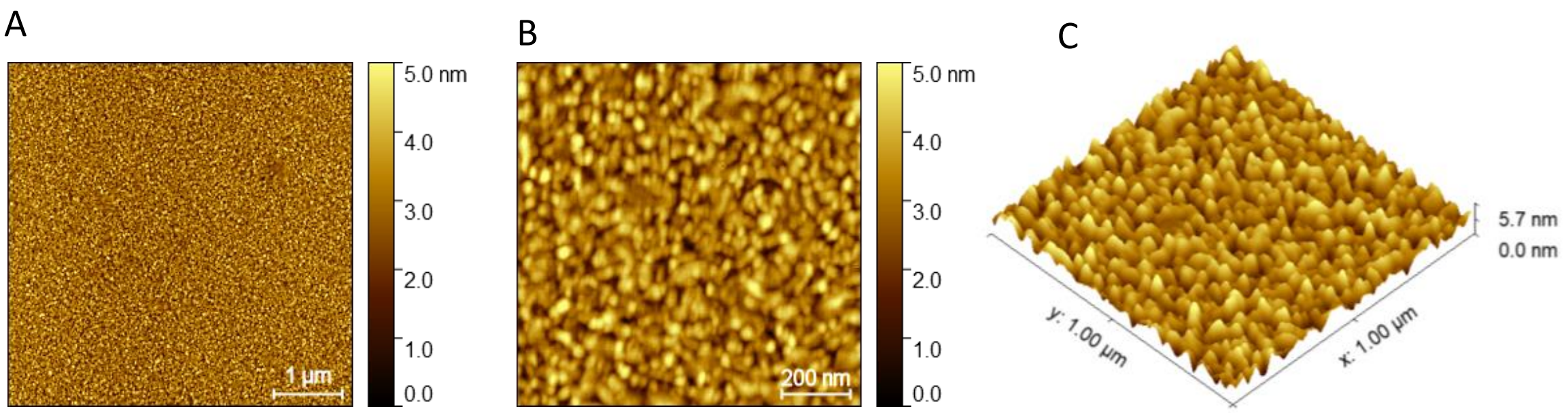

**Figure S1.** AFM topographic images of a 10 nm carbon film deposited on a flat Au substrate show a conformal coating with dense and uniform coverage, exhibiting a surface *rms* roughness of 0.5 ± 0.05 nm. (**A**) Wide-area scan ($5 \times 5\ \mu m^2$), (**B**) high-resolution image ($1 \times 1\ \mu m^2$), and (**C**) 3D representation of image (B).

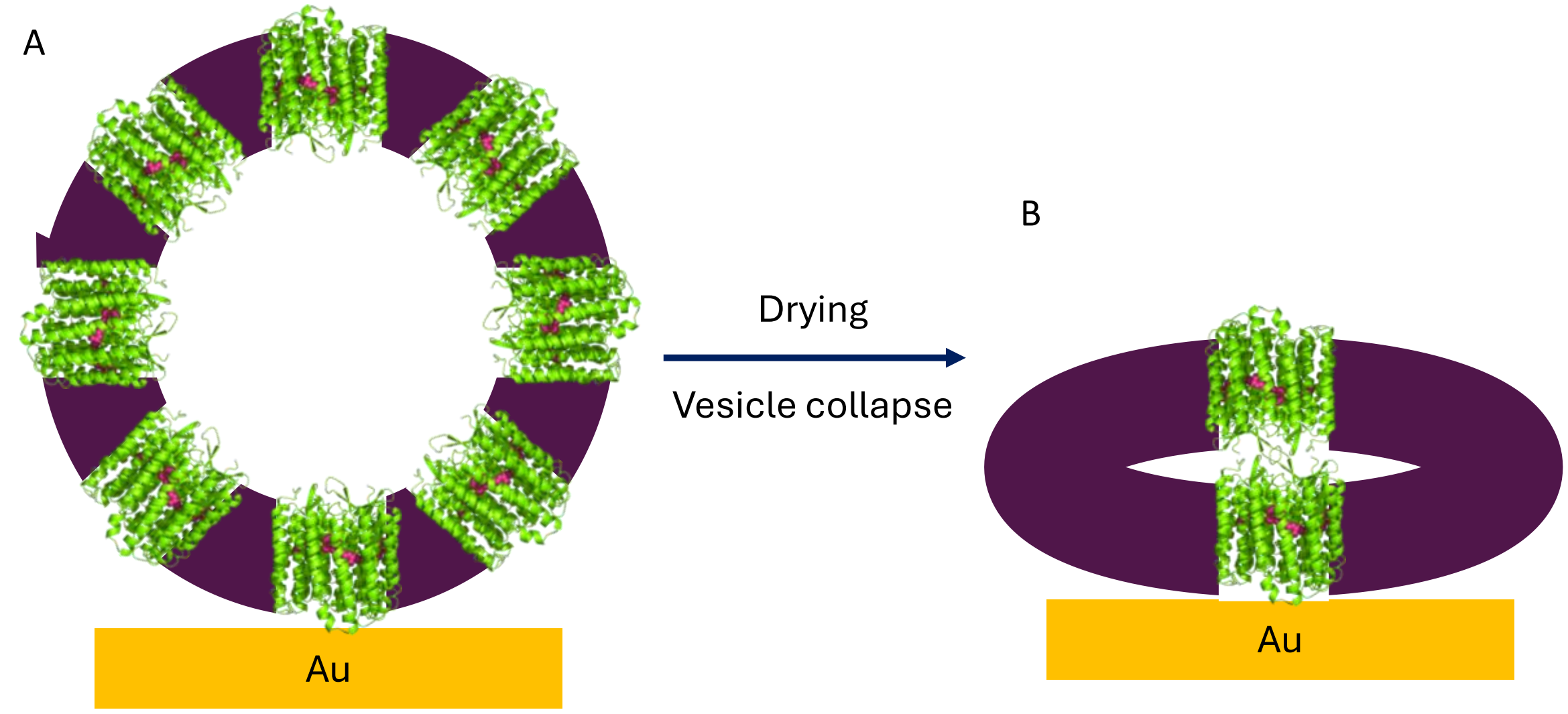

**Figure S2.** OTG-treated bacteriorhodopsin (bR) vesicle, electrostatically immobilized on a linker (cysteamine, Cys)-functionalized gold (Au) substrate. The vesicle is doughnut-like shaped with embedded bR trimers (green ribbon clusters; PDB ID: 1BRR) within the native lipid matrix (purple; simplified representation). The negatively charged outer vesicle surface enables electrostatic attachment to the positively charged linker layer, while the top-facing surface remains predominantly non-polar. (**A**) Single bR vesicle immobilized on a Au/Cys surface in buffer solution. (**B**) Upon drying, the immobilized vesicle collapses, forming a stacked bilayer structure composed of successive trimeric bR arrays (these need not stack exactly, but the differences in polar character make that the energetically favorable way). The resulting single bR bilayer (bR-SL), modeled using crystallographic dimensions, closely matches the experimentally determined protein layer thickness.

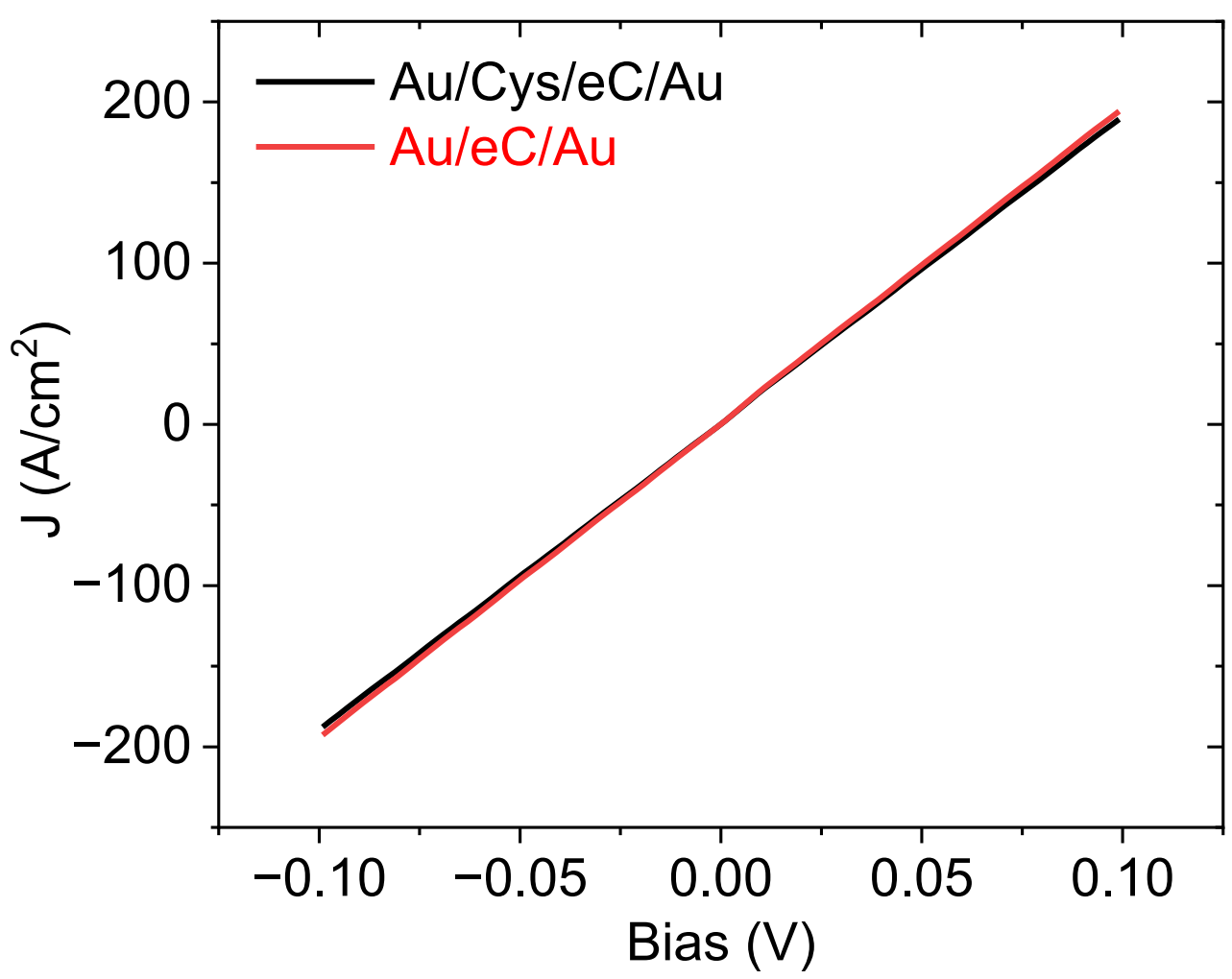

**Figure S3.** Representative current density ($J$)–voltage ($V$) characteristics of the shorted device with a bare electrode and a linker-coated device, as indicated in the figure legend, measured at room temperature (293 ± 2 K).

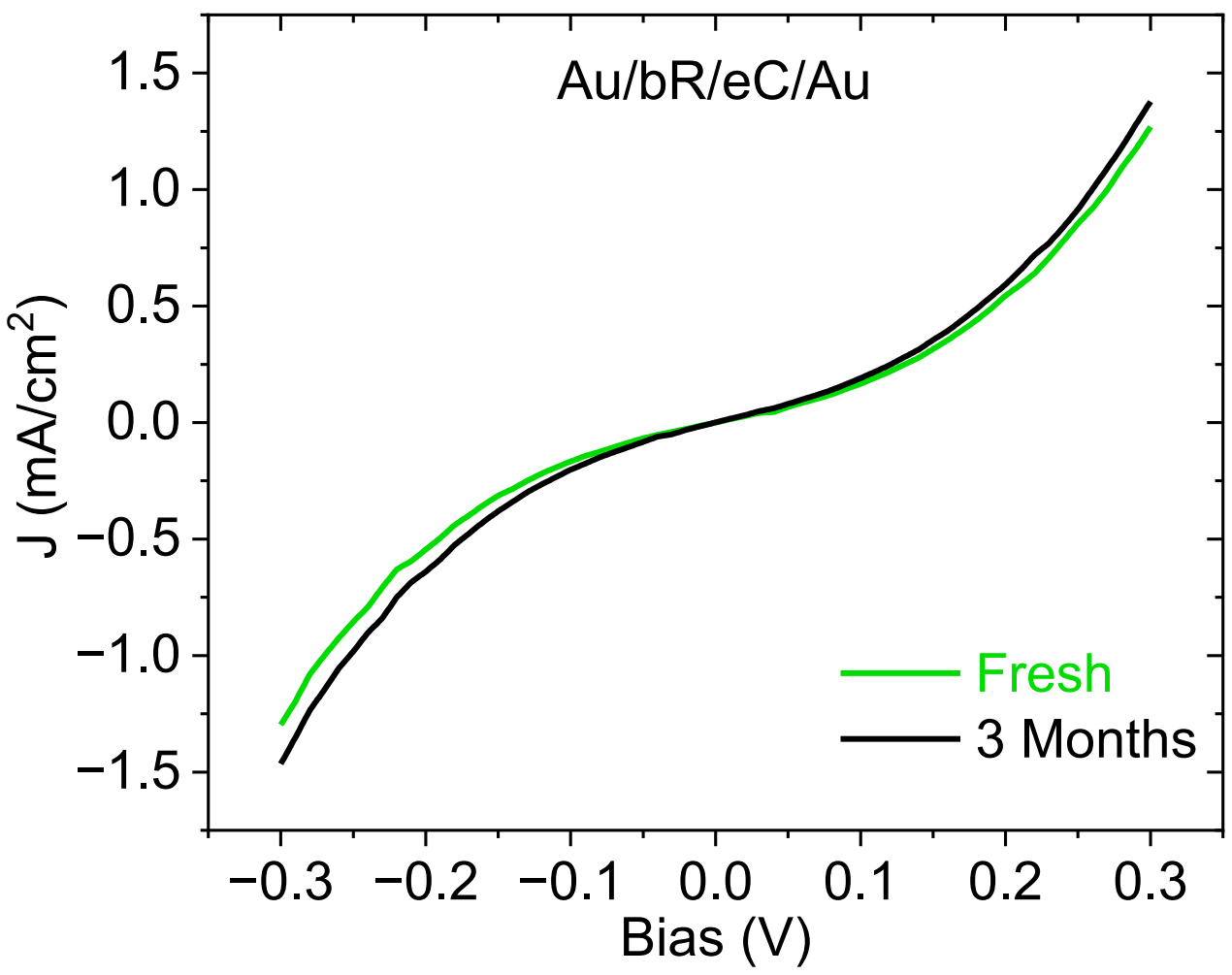

**Figure S4.** Representative current density ($J$)–voltage ($V$) characteristics of transport-active bR-SBL junctions measured over time, from freshly prepared to 3-month-old devices, showing negligible variation in current response (note the linear current density scale). Measurements were conducted at room temperature (293 ± 2 K).

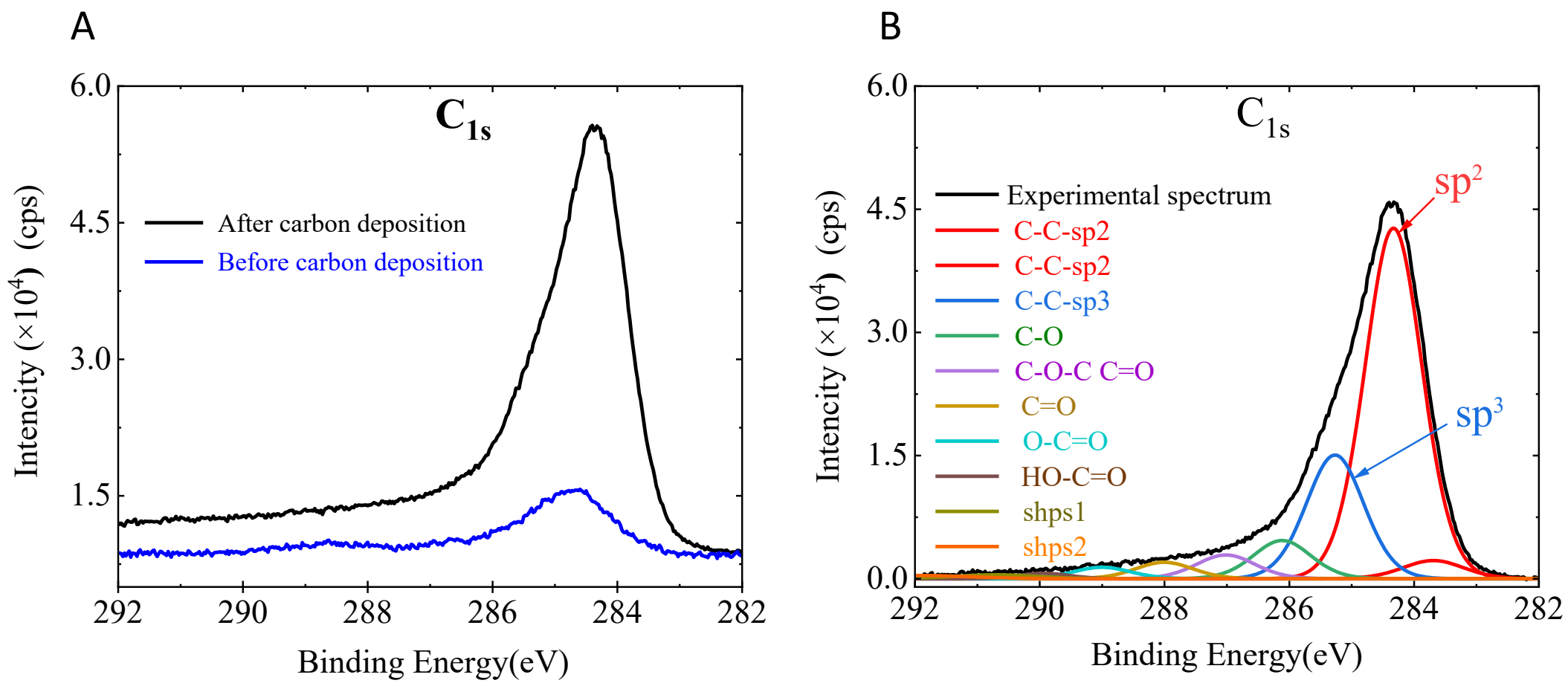

**Figure S5.** (**A**) High-resolution $C_{1S}$ XPS spectra recorded before (blue) and after (black) eC deposition. (**B**) Deconvoluted $C_{1S}$ spectrum obtained after deposition, highlighting the $sp^2$, $sp^3$, and other carbon-bonded functional fragment components. The corresponding relative atomic percentages are listed in **Table S1**.

| Deconvoluted $C_{1S}$ spectrum analysis | | | | | | | | |
|---|---|---|---|---|---|---|---|---|
| Carbon-bonded functional fragment components | C-C $sp^2$ | C-C $sp^3$ | C-O | C-O-C C=O | C=O | O-C=O | HO-C=O | Cshake-up |
| **Atomic %** | 61.42 | 20.76 | 6.4 | 4.07 | 2.74 | 1.91 | 0.81 | 1.89 |

**Table S1:** Relative composition of chemical type of carbon (-bonded functional fragment components) in our experimentally deposited carbon film in atomic percentage extracted from the area under the deconvoluted spectrum of the components of $C_{1S}$.

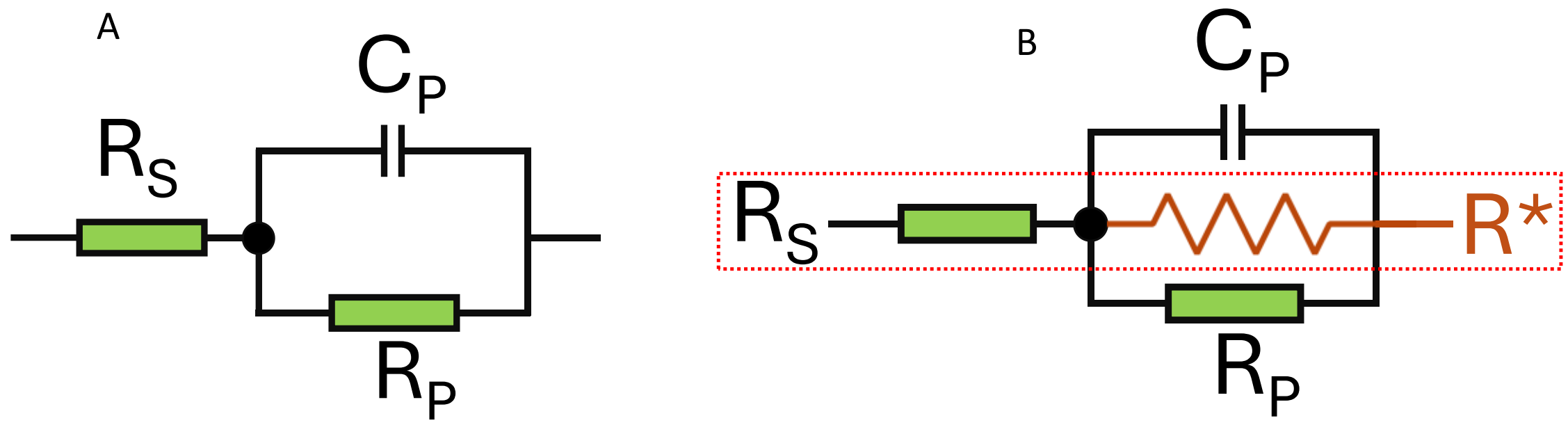

**Figure R6.** (**A**) The equivalent circuit for a transport active protein junction; (**B**) equivalent circuit for a filamentous-shorted junction, with the effective reduced circuit encircled inside the red dotted box.

**References**

[1] C. D. Wagner, https://nvlpubs.nist.gov/nistpubs/Legacy/TN/nbstechnicalnote1289.pdf
[2] A. Morteza Najarian, B. Szeto, U. M. Tefashe, R. L. McCreery, *ACS Nano* **2016**, *10*, 8918.
[3] S. Bera, S. Govinda, J. A. Fereiro, I. Pecht, M. Sheves, D. Cahen, *Langmuir* **2023**, *39*, 1394.
[4] S. Bera, J. A. Fereiro, S. K. Saxena, D. Chryssikos, K. Majhi, T. Bendikov, L. Sepunaru, D. Ehre, M. Tornow, I. Pecht, A. Vilan, M. Sheves, D. Cahen, *J. Am. Chem. Soc.* **2023**, *145*, 24820.
[5] S. Bera, A. Vilan, S. Das, I. Pecht, D. Ehre, M. Sheves, D. Cahen, *Advanced Materials* ***2025,*** e07654. https://doi.org/10.1002/smll.202506560